\documentstyle[proceedings,numreferences]{crckapb}

\makeindex
\begin{opening}
\title{QUANTUM MECHANICS AS AN EXOTIC PROBABILITY THEORY}

\author{SAUL YOUSSEF}
\institute{Supercomputer Computations Research Institute\\
           Florida State University\\
           Tallahassee, FL 32306-4052\\
           youssef@scri.fsu.edu\footnote{Presented at the Workshop on Maximum
Entropy and Bayesian Methods,
St.John's College, Santa Fe, New Mexico, August, 1995.}
}
\end{opening}
\begin{document}
\begin{abstract}
    Recent results suggest that quantum mechanical phenomena may
be interpreted as a failure of standard probability theory and
may be described by a Bayesian complex probability theory.
\end{abstract}
\index{complex probability theory}

   There is more to probability theory than proving theorems
in a particular mathematical system.  One is also in a position to
make predictions about real physical systems by adding extra assumptions
to the standard axioms.
Such predictions are necessarily
subject to experimental test, and, to the extent that one believes
in the extra assumptions, such tests may be interpreted as testing
the correctness of probability theory itself.  Now this may already
seem like an
odd point of view, especially here, since this conference series itself
provides
a most impressive record of success for probability theory in a vast
array of situations with no indication of a problem --
so why is there any reason to doubt probability theory?
Here I think that there is a historical effect:
probability theory may actually be failing all the time, it's just that the
situations where a failure occurs are called ``quantum mechanical
phenomena'' and thus appear in physics conferences instead of
in probability theory conferences.
This suggests that perhaps
there is something wrong with probability theory after all, and that
this may be where quantum mechanical effects come from.  Let's adopt this
point of view and see where it leads
\cite{sy1,sy2,sy3}.

   An obvious place to test our new point of view is the
two--slit experiment where, as everyone knows,
the fact that an interference pattern is observed even if one particle
is sent through at a time, forces us to conclude that it is not true
that a particular particle either goes through {\it slit $\#1$} or through
{\it slit $\#2$}; in general, then, a particle cannot be said to follow
a path through space.  This is the ``wave--particle duality,'' the basic
effect in quantum mechanics.  Notice, however, that
from our new point of view, the standard argument has a hole in
it due to it's essential reliance on probability theory.
For a position $x$ on the screen where there is a dip in the interference
pattern, one reaches a contradiction by noticing that
$$
{\rm P}(x) = {\rm P}(x\ via\ slit \#1)+{\rm P}(x\ via\ slit \#2)\geq {\rm P}(x\
via\ slit \#1)\eqno(1)
$$
means that opening the second slit should not cause the probability
to arrive at $x$ to decrease.  But if we are willing to modify
probability theory, then the standard argument and it's surprising conclusions
do not necessarily follow.  In fact it is clear that in order to escape the
standard conclusions, a modified probability theory must provide
a way for probabilities to cancel each other and so an obvious first guess is
to allow probabilities to be complex numbers.  Here, of course, the argument
grinds to a halt for a frequentist since frequencies are not the complex
numbers.  However, as Bayesians, we are not completely out of options because
for us, probabilities start out only as (real and non--negative)
measurements of ``likelihood'' where the frequency meaning for this
likelihood is derived after the fact.  Similarly, we might consider a
complex ``likelihood'' and see if a frequency meaning can
be found for this as well.  In fact, the simplest thing to do is to take Cox's
assumptions\cite{cox} and just drop the restriction that probabilities be real
and non--negative.  In this case, it turns out that Cox's entire
argument follows as before and one ends up with ``complex probability theory''
having the same form as the probability theory that you're used to except that
probabilities are complex.  For any propositions $a$, $b$ and $c$,
$$
(a\rightarrow b\wedge c) = (a\rightarrow b) (a\wedge b\rightarrow c)\eqno(I.a)
$$
$$
(a\rightarrow b) + (a\rightarrow\neg b) = 1\eqno(I.b)
$$
$$
(a\rightarrow false) = 0\eqno(I.c)
$$
where I have written the complex probability that proposition $b$ is true given
that proposition $a$ is known as ``$(a\rightarrow b)$'' (to be read: ``$a$ goes
to $b$'') reserving
the more familiar notation $P(b|a)$ for standard $[0,1]$ probabilities.

  Given our complex probability theory we would like to continue in parallel
with the Bayesian development
and construct a frequency meaning for complex probabilities.  Recall that for
standard probability, this works by supposing that the probability of something
is $p$ and considering $N$ copies of that situation with $f=n/N$ successes.
Using the central limit theorem $f$ is asymptotically gaussian
with mean $p$.  Then, since the probability for $f$ to
be in any interval not containing $p$ can be made arbitrarily small by
increasing
$N$, this fixes the frequency meaning for $p$.  Essentially, the frequency
meaning
of $p$ rests on the extra assumption that an arbitrarily small probability for
$f$ to be in some interval means that $f$ in fact will never be observed to be
in that interval in a real experiment.  The situation is not quite so simple in
complex probability theory because a zero complex probability does not in
general
mean that the corresponding event will never happen.  However, we can proceed
by
assuming that this extra condition is true for a special set of propositions
$U$ called
the ``state space.'' Let's also assume that this $U$ satisfies the following
for $x,y\in U$, propositions $a,b$, time $t\leq t'\leq t''$:
$$
x_t\wedge y_t=false\ {\rm if}\ x\neq y\eqno(II.a)
$$
$$
(a_t\rightarrow b_{t''}) = (a_t\rightarrow U_{t'}\wedge b_{t''})\eqno(II.b)
$$
$$
(a_t\wedge x_{t'}\rightarrow b_{t''})=(x_{t'}\rightarrow b_{t''})\eqno(II.c)
$$
where subscripts denote time, as in ``$a_t$'' meaning ``$a$ is true at time
$t$,''
and where a set of propositions with a subscript denotes the $or$ of each
element with the same subscript, as in $U_t=\vee_{x\in U}x_t$.
These are just Markovian style axioms intuitively corresponding to
``the system has a state.'' Roughly, the system cannot be in two different
states at the same time (II.a), the system is in some state at each
intermediate
time (II.b) and the knowledge that a system is at some point in the state space
makes all previous knowledge irrelevant (II.c).  We assume
that $U$ is a measure space with
$(a_t\rightarrow U_{t'})=\int_{x\in U}(a_{t}\rightarrow x_{t'})$.  Note the
clash
of terminology where the Hilbert space of standard quantum mechanics is
sometimes also
called the ``state space.''  Here $U$ is only a measure
space of propositions.

Given I and II, one can repeat the standard
argument for the expression
$$
Prob(a_t,b_{t'}) =
  {{\int_{x\in U} |a_t\rightarrow b_{t'}\wedge x_{t'}|^2 }\over
   { \int_{x\in U}   |a_t\rightarrow x_{t'}|^2 }}
$$
which predicts the frequency that $b$ is found to be true at time $t'$ given
that $a$ is known at a previous time $t$.  Although {\it Prob} as defined
is able to predict the frequencies for outcomes of any experiment,
it fails to extend to propositions involving mixed times
(e.g. $b_{t'}\wedge c_{t''}$, $t''>t'$).  This is an interesting point
because it is exactly this failure that allows complex probability
theory to escape Bell's theorem\cite{sy3}.  Also, although I don't have
a sharp result, it is seems likely that this effect disappears
in a classical limit, thus explaining why standard probability
theory works in the classical domain.

   Immediate consequences of axioms I.a--I.c and II.a--II.c
include facts familiar from probability theory
such as $(a\rightarrow true) = 1$,
$(a\rightarrow b\vee c) = (a\rightarrow b) + (a\rightarrow c) - (a\rightarrow
b\wedge c)$,
and if $(a\rightarrow b)\neq 0$, then
$(a\wedge b\rightarrow c)=(a\rightarrow c)
(a\wedge c\rightarrow b)/(a\rightarrow b)$ {\it (Bayes Theorem)}.
Following standard probability theory, propositions $a$ and $b$ are said
to be {\it independent} if $(q\wedge a\rightarrow b)=(q\rightarrow b)$ for all
$q$
and, just as in standard probability theory, ``locality'' enters via
assumptions
of independence.  For instance, if experiments $e_1$ and $e_2$ have possible
results
$r_1$ and $r_2$ respectively, then the assumptions that
$\{r_1,r_2\}$, $\{e_1,r_2\}$, and $\{e_2,r_1\}$ are independent imply
$$
(e_1\wedge e_2\rightarrow r_1\wedge r_2) = (e_1\rightarrow r_1)(e_2\rightarrow
r_2)
$$
as one would expect from, for example, two experiments which have nothing to do
with
each other.  Other simple consequences of the axioms are described in
references 1-4
including
\begin{itemize}
\item{The Path Integral}
\item{The Superposition Principle}
\item{The Expansion Postulate}
\item{The Schr\"odinger/Klein--Gordon Equations for $U=R^d$}
\end{itemize}
where the standard wavefunction $\Psi(x,t)$ is proportional to the complex
probability
$$
(Everything\ that\ you\ know\ about\ the\ system\rightarrow x_t)
$$
making the Bayesian status of the wavefunction obvious.  In particular, the
same system may be described by different wavefunctions depending upon what is
known
and such wavefunctions can clearly not be ``the state of the system'' in
any reasonable sense.

   To get a feeling for how things work, let's consider a typical
interferometer
as shown in figure 1.
\begin{figure}
\vglue4in
\includegraphics{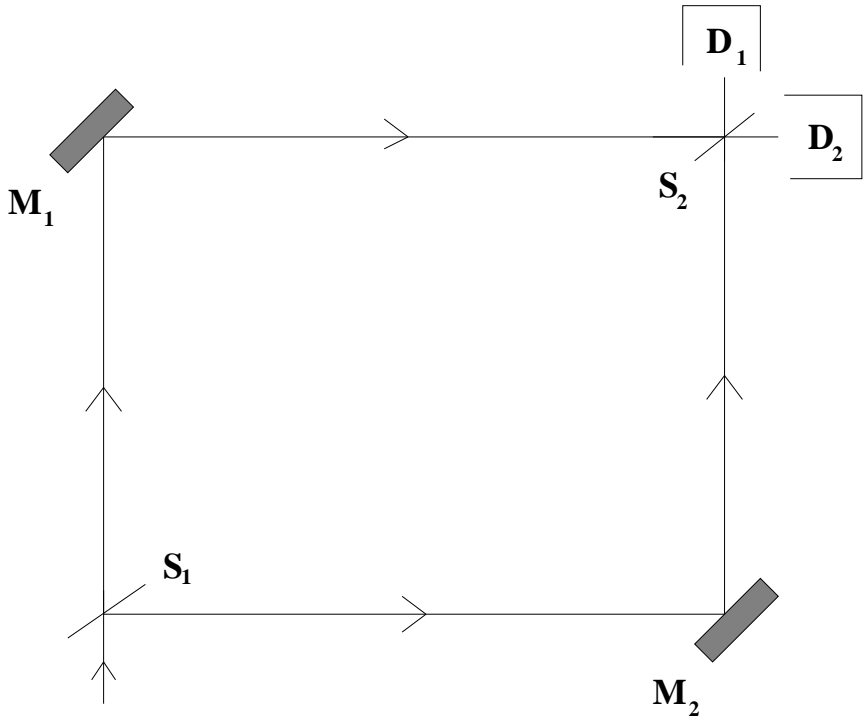}
\caption{A simple interferometer where a particle enters as indicated and
encounters a beam splitter ($S_1$), a mirror ($M_1$ or $M_2$), a second
beam splitter ($S_2$) ending up either in detector $D_1$ or in detector $D_2$.}
\end{figure}
Particles enter the device one at a time, pass through
a beam splitter, hit one of two mirrors and pass through a second beam splitter
ending up either in detector $D_1$ or in $D_2$.  Although it looks perfectly
possible
for a particle to end up in $D_2$, experimentally we mysteriously find that
this
never happens.  All the particles register in $D_1$.
In standard quantum theory, one describes this situation
by saying that there are two paths for a particle to go from the source to
$D_2$.  The amplitude for these paths have opposite signs and since the
probability
is the square of the total amplitude, this explains why particles never enter
$D_2$.
Now let's consider a modified situation (figure 2) where a device is
attached to one of the mirrors which is able to detect if the mirror was
struck by the particle.
\begin{figure}
\vglue4in
\includegraphics{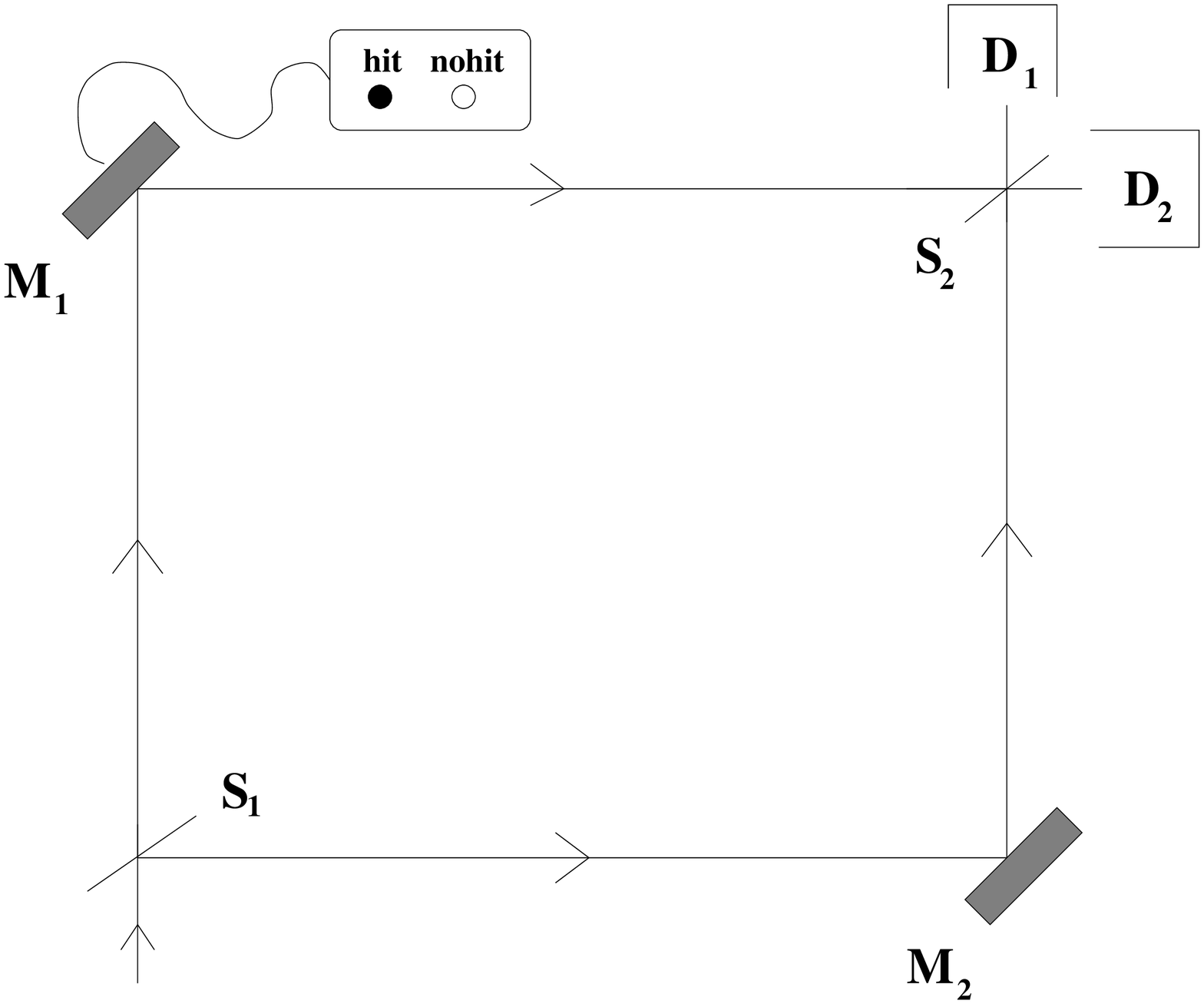}
\caption{A interferometer similar to the one shown in figure 1.  The device
is attached to the mirror $M_1$ records whether $M_1$ was struck
by the particle or not.}
\end{figure}
After each particle passes through, the device either indicates ``hit'' or ``no
hit.''
Experimentally, we find that the results are now different with about
half of the particles ending up in detector $D_2$.  How can we explain this?
In standard quantum mechanics there are still two paths for a particle to
reach $D_2$ as before.  However, since we can now tell which path was taken
by the particle by inspecting the hit/no--hit device, the amplitudes for the
two paths no longer interfere.  This is a special case of the general
principle:
\begin{itemize}
\item{{\it Paths interfere only if there is no
way of knowing which path was taken, even in principle.}}
\end{itemize}

\noindent This is a rather mysterious statement since it suggests that whether
or not you can deduce
which path was taken somehow affects the behavior of the particle.  Also, it is
not clear what ``even in principle'' means here or what happens
if, for example, the hit/no--hit device works with, say, 99\% efficiency.
Even so, the basic prediction of this principle is correct and this raises
the question of whether there is an analogous principle in complex probability
theory and whether the predictions are the same.
You can indeed easily deduce from the axioms and the definition of $Prob$ that
\begin{itemize}
\item{{\it Paths interfere if and only if they end at the same point in U.}}
\end{itemize}

\noindent and this appears to give the same predictions as the more mysterious
sounding
standard quantum mechanical principle.  For instance, within complex
probability
theory, the situation of figure 1 could be described by a state space $R^3$
(assuming that the particle is spinless) in which case the complex probability
to arrive at $D_2$ is the sum of the complex probabilities for two paths, which
also cancel, just as in standard quantum mechanics.  In the situation shown
in figure 2, one simply
notes that the state space $R^3$ is evidently no longer sufficient to describe
the
system.  If one extends the state space to, say, $U=R^3\times\{hit,nohit\}$,
then
the interference is lost because the two paths for reaching $D_2$ now end at
different points in $U$.  There is also a continuum between this result,
where the hit/no--hit device is assumed to work perfectly, and a situation
where
the device works so badly that the propositions ``hit'' and ``no--hit'' are
independent
of which path the particle is taking.
In this case,
you can easily show that the original effect is restored\cite{sy2}.

   Of course, standard quantum mechanics is perfectly capable of handling a
situation
like that of figure 2.  It's just that a rigorous
treatment of the problem with a Hilbert space including the hit/no--hit device
where the state vector evolves under the action of a Hamiltonian would be
rather
difficult, especially considering the simplicity of the answer.
This helps to explain the popularity of ``which path'' style arguments
in spite of their ambiguity.  Here complex probability theory has the advantage
that simple assumptions about a system can rigorously be encorporated without
having to decide what the assumptions mean in terms, for example, of solutions
to the
Schr\"odinger equation.
A rigorous treatment of these problems within
quantum mechanics would also have to address the issue of whether the initial
state is ``mixed'' or not since not all situations in quantum mechanics can be
described by a vector in a Hilbert space, some require ``statistical mixtures''
of
vectors in a Hilbert space.  This provides an interesting test for complex
probability
theory.  Since ordinary ``statistical mixtures'' are no longer available to us,
situations
requiring ``mixed states'' had better be handled within the existing axioms.
These situations appear to indeed be handled quite smoothly and naturally
within the complex probability theory described here\cite{sy2}.

  To take an example with more detailed predictions, consider a single scalar
particle with $U=R^d$ and consider a sequence of propositions
$x_o, x_1,\dots, x_n$ in $U=R^d$ where each $x_j$ implicitly has a time
subscript
$t_j$ with $t_{j+1}-t_j=\tau>0$.
As always, $(x_o\rightarrow x_n)$ is
given by the ``path integral''
$$
(x_o\rightarrow x_n) = \int_{x_1}\dots\int_{x_{n-1}}(x_o\rightarrow
x_1)(x_1\rightarrow x_2)
\dots(x_{n-1}\rightarrow x_n).
$$
Of course, there is a path integral in standard quantum mechanics as
well\cite{feynman}
where one proceeds to dynamics by assuming that the amplitude for a path
is proportional to $e^{i A}$ where the ``action'' $A$ is the time integral
of a classical Lagrangian for the system.  Here we can avoid these extra
assumptions
by repeating the same argument within each $\tau$ sized
interval making $n$ sub--path integrals (figure 3)
with time step $\epsilon=\tau/N$.
\begin{figure}
\vglue2.5in
\includegraphics{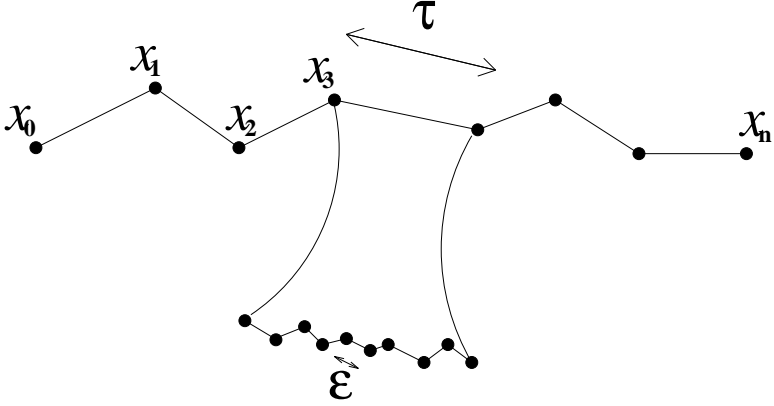}
\caption{For $x_0,x_n\in U$, the complex probability $(x_0\rightarrow x_n)$ is
equal to
a ``path integral'' over $x_1,x_2,\dots,x_{n-1}$ with time step $\tau$.  The
argument
can be repeated making $n$ sub--path integrals with time step $\epsilon$.}
\end{figure}
By letting both $\tau$ and $\epsilon$ go to zero, one
can extract a central--limit--theorem--like result where for small $|z|$ and
small
$\tau$, $(x_t\rightarrow (x+z)_{t+\tau})$ is given by
$$
{1\over (2\pi\tau)^{{d\over 2}}\sqrt{{\rm det}[\nu]}}
{\rm exp}(-\tau[{1\over 2}({z_j\over \tau}-\nu_j)W_{jk}({z_k\over \tau}-\nu_k)
+ \nu_0])
%{\rm exp}(-\tau\left[{1\over 2}({z_j\over \tau}-\nu_j)W_{jk}({z_k\over
%%\tau}-\nu_k) + \nu_0\right])
$$
where $\nu_o(x)$, $\nu_j(x)$, $\nu_{jk}(x)$ and $W_{jk}(x)$ are
moments of $\mu(x,z,\tau)\equiv(x_t\rightarrow (x+z)_{t+\tau})$
defined by
\begin{description}
\item{$\nu_0(x) = \int_{z\in U} \mu_{\tau}(x,z,0)$}
\item{$\nu_j(x) = \int_{z\in U} \mu_{\tau}(x,z,0)z_j$}
\item{$\nu_{jk}(x) = \int_{z\in U} \mu_{\tau}(x,z,0)z_j z_k$}
\end{description}
with $W_{jk} = M_{jl}M^{T}_{lk}\omega_l$ where $M$ is the matrix which
diagonalizes $\nu_{jk}$
such that $M^{T}_{lj}\nu_{jk}M_{km} = \delta_{lm}/\omega_l$.  With velocity
$v_j$ given by the limit
of $z_j/\tau$, the above
propagator is equivalent to the
Lagrangian
$$
{\cal L}(x,v) = {i\over 2}(v_j-\nu_j)W_{jk}(v_k-\nu_k) - i\nu_0
$$
where we recognize $\nu_j(x)$ and $\nu_o(x)$ as the electromagnetic fields
and where $W_{jk}(x)$ contains the particle mass and space--time metric.
Notice that we have not assumed Lorentz or gauge invariance to get this
result.  The claim is that this is the only Lagrangian consistent with
the state space $R^d$.

   Since our complex probability theory is both ``realistic'' in the sense of
assuming that a particle does come through one slit or the other in the two
slit experiment (II.b) and local in the sense of accepting locality assumptions
as assumptions of ``complex statistical independence,'' you might think that we
would run afoul of Bell's theorem or other more recent limitations on local
realistic
theories.  As I've already mentioned, Bell's result does not follow in complex
probability theory.  This means that although Bell's result has almost
universally
been interpreted
as ruling out local realistic theories, from our more general point of view, it
forces a choice between local realism and standard probability theory.
In fact, Bell's result can be interpreted as another little hint that
there is something wrong with probability theory.
Besides Bell, there are a large number of more recent limitations
on local realistic theories, each of which provides a test of the complex
probability formulation.
Details are available only for three representative
results of this type (including Bell) where complex probability theory appears
not to be excluded\cite{sy3}.

   You might expect that if quantum mechanical phenomena can be described
by complex probability theory, the Bayesian view might help in understanding
some of the long standing semi--paradoxical measurement and observer
questions in quantum mechanics.
Here, it's helpful to first think about a purely classical experiment where
a single coin is flipped and then uncovered, revealing that it landed
``heads.''
{}From the Bayesian point of view, of course, the situation before
the observation could be described by the distribution $(1/2,1/2)$ and
after observing heads our description would be adjusted to $(1,0)$.
The problem is, what would you say to a student who then asks:
\begin{itemize}
\item{{\it Yes, but what causes $(1/2,1/2)$ to
evolve into $(1,0)$?  How does it happen?}}
\end{itemize}
Here we recognize a victim
of a severe form of ``Mind Projection Fallacy''\cite{jaynes89}
where the person asking this question has confused what they
know about the system with the system itself.
With the Bayesian view of
complex probabilities, it is clear that this same mistake is possible
in quantum mechanics as well,
where one would now be mistaking a wavefunction for the state of the system.
This very view, however, is the standard picture of
quantum mechanics and so it is hardly surprising that similar mysteries arise.
This view is also implicit in questions such as
\begin{itemize}
\item{Can the wavefunction be measured?}
\item{What is the source of the non--local effects in EPR?}
\item{Can macroscopic superpositions be created?}
\item{Is the Universe in a pure state?}
\end{itemize}
Although these are active research questions, it seems inescapable to me that
if quantum phenomena are correctly described by a Bayesian probability
theory, then all of these questions have trivial answers and they all
ultimately
fall into the same category as the student's question about coin flipping
experiments.

   For this audience, I hardly need to point out that the ideas that we
have discussed here may not only clarify the meaning of quantum mechanics,
but may also lead to ways of
improving quantum mechanical calculations using prior knowledge in the
same sense that prior knowledge is used to improve probability calculations
in Bayesian Inference.


\begin{thebibliography}{}
\bibitem[\protect\citeauthoryear{Youssef}{1991}]{sy1}
S.Youssef, {\it Mod.Phys.Lett.}\ {\bf A6}, 225 (1991).
\bibitem[\protect\citeauthoryear{Youssef}{1994}]{sy2}
S.Youssef, {\it Mod.Phys.Lett.}\ {\bf A9}, 2571 (1994).
\bibitem[\protect\citeauthoryear{Youssef}{1995}]{sy3}
S.Youssef, {\it Phys.Lett.}\ {\bf A204}, 181(1995).
\bibitem[\protect\citeauthoryear{Cox}{1946}]{cox}
R.T.Cox, {\it Am.J.Phys.}\ {\bf 14}, 1(1946).
\bibitem[\protect\citeauthoryear{Feynman, Hibbs}]{feynman}
R.P.Feynman and H.R.Hibbs, {\it Quantum Mechanics and Path Integrals}
(McGraw--Hill, 1965).
\bibitem[\protect\citeauthoryear{Jaynes}{1989}]{jaynes89}
E.T.Jaynes, in {\it Maximum Entropy and Bayesian Methods}, ed.
J.Skilling(Kluwer, 1989).
\end{thebibliography}
\end{document}